\documentclass[aps,prd,amsfonts,amssymb,amsmath,showpacs,letterpaper,superscriptaddress,nofootinbib,altaffilletter]{revtex4-1}

\usepackage{tabularx}
\usepackage[final]{hyperref}
\usepackage[dvipsnames]{xcolor}
\usepackage{listings}
\usepackage[toc,page]{appendix}
\usepackage{docmute}
\usepackage{diagbox}
\usepackage{hyperref}

\usepackage{color}
\usepackage{graphicx}
\usepackage{latexsym}
\usepackage{float}
\usepackage{amsmath}
\usepackage{amssymb}
\usepackage{multirow}
\usepackage{ifthen}
\usepackage{slashed}
\usepackage{subfigure}
\usepackage{lineno}

\newcommand{\makevisible}[1]{\textcolor{red}{#1}}
\newcommand{\switch}[1]{%
  \ifthenelse{\equal{#1}{0}}{\renewcommand{\makevisible}[1]{}}{}}
\switch{1} 
\def\version$#1,v #2 #3${#2}

\begin{document}


\title{Improving the background of gravitational-wave searches for core collapse supernovae: A machine learning approach}

\author{M.\ Cavagli\`a}
\affiliation{Missouri University of Science and Technology, 1315 N.\ Pine St., Rolla MO 65409, USA}
\author{S.\ Gaudio}
\author{T.\ Hansen}
\affiliation{Embry-Riddle Aeronautical University, Prescott AZ, USA}
\author{K.\ Staats}
\affiliation{Northwestern University, Evanston IL, USA}
\author{M.\ Szczepa\'nczyk}
\affiliation{University of Florida, Gainesville FL, USA}
\author{M.\ Zanolin}
\affiliation{Embry-Riddle Aeronautical University, Prescott AZ, USA}

\date[\relax]{Dated:\today}

\begin{abstract}

\noindent Based on the prior O1-O2 observing runs, about 30\% of the data collected by Advanced LIGO and Virgo in the next observing runs are expected to be single-interferometer data, i.e., they will be collected at times when only one detector in the network is operating in observing mode. Searches for gravitational wave signals from supernova events do not rely on matched filtering techniques because of the stochastic nature of the signals. If a Galactic supernova occurs during single-interferometer times, separation of its unmodelled gravitational-wave signal from noise will be even more difficult due to lack of coherence between detectors. We present a novel machine learning method to perform single-interferometer supernova searches based on the standard LIGO-Virgo coherent WaveBurst pipeline. We show that the method may be used to discriminate Galactic gravitational-wave supernova signals from noise transients, decrease the false alarm rate of the search, and improve the supernova detection reach of the detectors.

\end{abstract}


\maketitle

\section{Introduction}

A principal challenge in detecting gravitational waves (GWs) is distinguishing astrophysical signals from instrumental or environmental noise triggers produced by non-linear couplings between the detector subsystems and/or their environment \cite{TheLIGOScientific:2017lwt,TheLIGOScientific:2016zmo,Aasi:2014mqd}.

If the theoretical GW signal is known, as in the case of binary coalescences \cite{2018arXiv181112907T}, triggers are generated with a matched-filter technique \cite{Allen:2005fk,Sathyaprakash:1991mt,Owen:1998dk}. In a multi-detector array, such as the current Advanced LIGO \cite{TheLIGOScientific:2014jea} and Advanced Virgo \cite{TheVirgo:2014hva} network, a transient GW signal should appear as a near-simultaneous trigger across all three detectors, the delay defined by the direction of travel of the gravitational wave and the associated light travel time.

The matched-filter technique cannot be used for unmodelled signals such as GWs emitted in Core-Collapse Supernovae (CCSNe) \cite{Abbott:2016tdt}. Despite recent
progress in numerical simulations, the dynamics of supernova explosions is not yet fully understood as the extremely complex physics of star collapse and the
computational cost required for accurate simulations make the treatment of CCSNe very challenging. Theoretical and computational improvements over the last few years
have allowed several teams to calculate some CCSN GW waveforms through different approximations and numerical schemes in two- and three-dimensional scenarios
\cite{2017arXiv170304633M,2017arXiv170107325Y,Kotake:2006aq,Powell2019}. The main time frequency features for slowly rotating progenitor stars are the
progressive increase of the dominant mode frequency and an occasional development of the constant-frequency Standing Accretion Shock Instability (SASI). For rapidly
rotating progenitors there is consensus in the simulation community for a strong broad band, temporally very compact (few tens of milliseconds) component, although
the later stages of GW production are still under discussion. Even as the pool of available waveforms evolves and more GW waveforms appear in the literature, these
main features seem to be common across the various families of numerical simulations. The waveforms used in our analysis capture these features as currently visible
in published waveforms. (For more examples, see Refs.\ \cite{Radice:2018usf,OConnor:2018tuw,Andresen:2018aom,Kuroda:2016bjd,Andresen:2016pdt,Muller:2011yi}.) In addition,
CCSNs are stochastic processes. Therefore, state-of-the-art waveforms are not yet sufficiently reliable to be used as templates in a matched-filter search and cover
the full search parameter space. 


GW signals from CCSNe are typically much weaker than GW signals from binary mergers. Because of this, and the stochastic nature of the signal, the background of GW searches from CCSNe is expected to be severely polluted by short duration noise transients which may mimick actual signals. The situation is even worse when coincident data from multiple detectors are not available. During the first and second LIGO-Virgo observing runs (O1 and O2) a significant portion of the LIGO-Virgo data ($\sim 33$\% in O1 and $\sim 30$\% in O2) were collected in ``single-interferometer'' mode, i.e., when only one detector in the network was operating in nominal observing configuration (see https://www.gw-openscience.org \cite{Vallisneri:2014vxa}). Even with improved detector reliability and duty cycle (70\%), it is expected that about 20\% of the data in the next observing runs, the LIGO-Virgo network will be single-interferometer data. Given the rarity of a Galactic CCSN~\cite{cappellaro:93,li:11,2013ApJ...778..164A}, extending the detector range and improving the search background is essential to maximize the chances of detection of a GW CCSN signal.

In this paper we present a novel technique based on a supervised Machine Learning (ML) algorithm \cite{KarooGP} which may be effectively employed in future LIGO-Virgo observing runs to reduce the background of single-interferometer data and achieve a 3$\sigma$ confidence level detection in GW searches for Galactic CCSN. Simpler approach without ML and using fewer GW signals was performed in Ref.~\cite{SzczepanczykThesis}. In our method we assume that the event time and the distance of the CCSN are known from neutrino and optical observations. The machine learning algorithm is first trained on off-source data to produce a lower background. The results are then applied to on-source windows around GW event candidates to increase the detection confidence. We train the algorithm on approximately 1.47 days of O1 data by injecting the set of waveforms that have been used in the latest LIGO-Virgo observing runs \cite{Abbott:2016tdt, O1O2CCSNe} to obtain CCSN detection upper limits at various fixed distances smaller than the distance to the Galactic center ($<10$\,kpc). The features of simulated and background triggers are extracted using the coherent-WaveBurst (cWB) pipeline \cite{Klimenko:2008fu,Klimenko:2015ypf} employed by LIGO and Virgo for unmodelled GW transient searches. The machine learning algorithm is then used to classify the triggers and remove the noise triggers.

\section{Analysis search pipeline}\label{cWB}

Our analysis utilizes coherent WaveBurst (cWB) analysis, a software pipeline widely used in the LIGO-Virgo Collaboration for the detection and reconstruction of unmodelled gravitational-wave (GW) events, e.g.~\cite{ligo_burst_s5y1:09, snsearch, abbott17}. At its core, cWB employs the constrained maximum likelihood ratio. The method combines data streams from ranking statistic $\rho$ that is the coherent network signal-to-noise ratio (SNR) of a GW signal detected in the network.


To perform its analysis, cWB requires data streams from at least two detectors, thereby reducing the population of the events associated with coincident noise between them. The principal challenge in establishing a detection with only one interferometer is that consistency constraints, that measure a degree of similarity of an event between different detectors, cannot be applied. Without consistency constraints, a population of loud noise glitches might persist in the data, and as a consequence, contribute to the reduction of statistical significance of GW induced candidate.


We propose a method that employs the cWB pipeline and machine learning algorithm to perform a single detector case analysis. First, we configure the first detector as an exact copy of the second detector such that the coherent analysis with cWB pipeline can be performed. Then, the statistical significance of the triggers is assessed with the False Alarm Probability (FAP) \cite{Abbott:2016tdt,O1O2CCSNe}:

\begin{equation}
 \mathrm{FAP} = 1 - e^{T_{on} \times \mathrm{FAR}}\,, 
\end{equation}

where FAR is the False Alarm Rate of the trigger and $T_{on}$ is the time period where we expect to find the signal (on-source window). Finally, the $\rho$ statistic becomes the SNR in the single detector case. 

The time of the collapsing core for a galactic CCSN is expected to be determined with an uncertainty of less than one second by the detection of a neutrino flux~\cite{PhysRevLett.103.031102}, so we set $T_{on}=2$\,s. We analyze 1.47 days of data from the Hanford detector during the O1 run, which would allow for a detection at $3\sigma$ significance level, corresponding to $\mathrm{FAP} \approx 2.7 \times 10^{-3}$.


The background analysis is performed across time-shifted data, thereby removing much of the potential for terrestrial noise or glitches to simulate a signal~\cite{Abbott:2016blz}. While time-shifted, pattern matching is used in the LIGO and Virgo searches, they are built on the assumption that two or three detectors are operational. In case of single detector analysis, the detector data streams cannot be time-shifted to produce the search background.

\begin{figure*}[t] 
  \begin{minipage}[c][][t]{0.495\textwidth}
    \vspace*{\fill}
    \flushleft
    \includegraphics[width=0.96\linewidth]{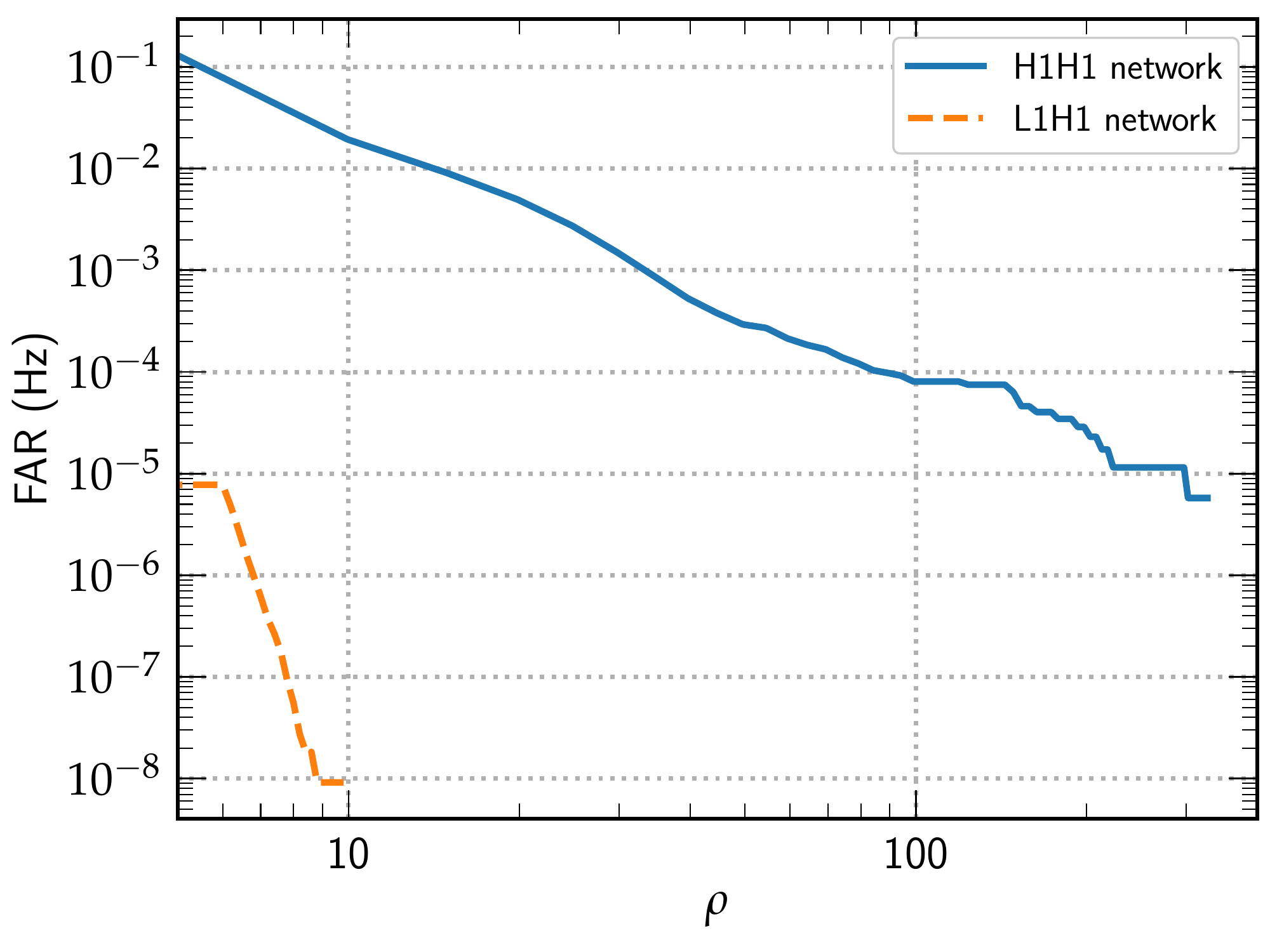}
  \end{minipage}%
  \begin{minipage}[c][][t]{0.495\textwidth}
    \vspace*{\fill}
    \flushright
    \includegraphics[width=0.96\linewidth]{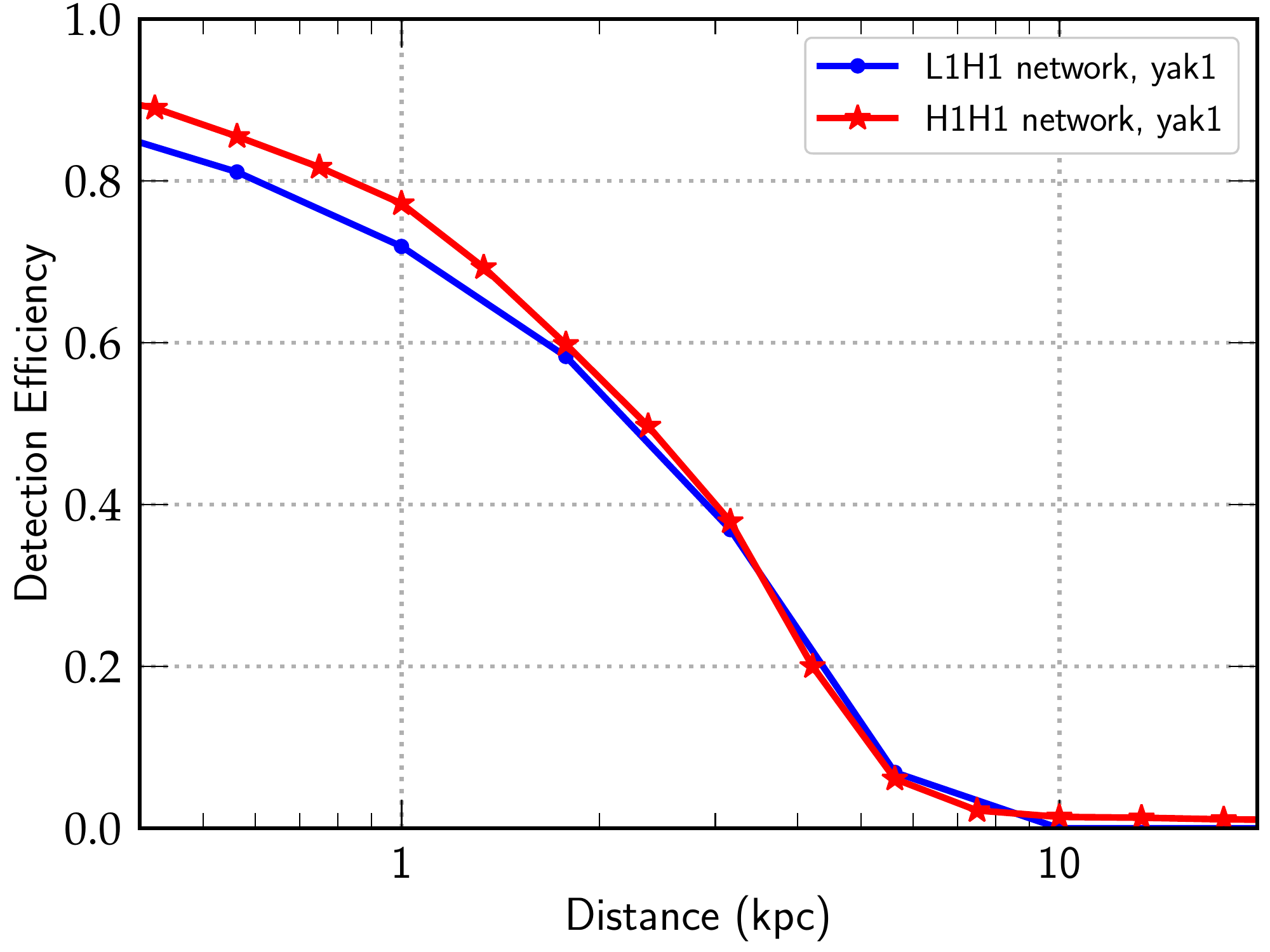}
  \end{minipage}
  \caption{\label{fig:sdc_initial}
  Comparison between background (left) and detection efficiency (right) for
  single detector case analysis and for L1H1 detector network.
  The background for L1H1 network is estimated using time-shifted data
  allowing to reach lower False Alarm Rate (FAR). Also, the consistency tests are not possible
  to perform in case of H1H1 network.
  On the other hand, the detection efficiencies in one and two detector analysis
  are comparable.
  }
\end{figure*}

The left panel of Fig.~\ref{fig:sdc_initial} shows the FAR of the background triggers for a H1H1 network and H1L1 network for comparison, where L1 and H1 denote the LIGO Livingston and LIGO Hanford interferometers. The FAR for the H1H1 network is orders of magnitudes larger than for a regular L1H1 network. Moreover, the noise triggers are much louder.

The cWB pipeline is also employed to analyze the detectability of GW signals generated from multidimensional simulations. The waveforms derived from
multidimensional simulations are added to the detector noise with amplitudes corresponding to an initial source distance of 10\,kpc. In our analysis, we
consider the set of waveforms that have been used in the latest LIGO-Virgo observing runs \cite{Abbott:2016tdt, O1O2CCSNe}. The pipeline first performs the
analysis to isolate the injected waveforms. The analysis is then performed several more times with the waveform amplitudes rescaled to a range of source
distances. For each waveform family, we inject the signals every 33\,s. As a result, we obtain a detection efficiency curve with respect to the source
distance.

Slowly or non-rotating massive stars are believed to constitute about 99\% of CCSNe \cite{Janka:2012wk}. (For the rates of different explosion
mechanisms see Ref.\ \cite{Gossan:2015xda} and references therein, as well as Refs.\ \cite{Li:2010kd,Cappellaro:1993ns}.) Among slowly or non-rotating models, we
choose for our analysis three distinct waveform families characterized by a neutrino-driven explosion mechanism \cite{muller,Ott:2012mr,Yakunin:2015wra}. From the
M\"uller family of three-dimensional numerical simulation waveforms \cite{muller}, we consider two waveforms calculated with a 15$M_\odot$ progenitor (\emph{mul1},
\emph{mul2}) and one waveform calculated with a 20$M_\odot$ progenitor (\emph{mul3}).  The Ott waveform is also generated by 3D simulations with a 27$M_\odot$. The
progenitor star is slowly rotating and convection is the dominating mechanism, leading to a less significant Standing Accretion Shock Instability (SASI)
\cite{Ott:2012mr} contribution to the waveform signal.

For the two-dimensional Yakunin model, we use four waveforms with progenitor masses of 12$M_\odot$, 15$M_\odot$, 20$M_\odot$ and 25$M_\odot$ \cite{Yakunin:2015wra}, respectively. We denote them with yak1, yak2, yak3 and yak4. The signal of these waveforms is generally very strong due to the fact that the axisymmetry of the (2D) system artificially increases the amplitude of the gravitational wave when compared to the other waveforms of this group.

We also examine the waveform models leading to explosion produced from rapidly rotating progenitors. For this case, we consider the Scheidegger
\cite{Scheidegger:2009va} and Dimmelmeier \cite{Dimmelmeier:2008iq} waveform families. The explosion mechanism for this second group is believed to be
magneto-hydrodynamically driven and the waveforms generally carry larger energies than in the non-rotating or slowly-rotating models. Scheidegger et
al.\cite{Scheidegger:2009va} calculate a large set of waveforms from three-dimensional simulations under various conditions. In our analysis, we choose three of
these waveforms with a 15$M_\odot$ progenitor star mass and different rotational speed, R1E1CA\_L (no rotation), R3E1AC\_L and R4E1FC\_L, which we denote sch1, sch2
and sch3, respectively. For the Dimmelmeier family \cite{Dimmelmeier:2008iq}, we choose three waveforms (dim1, dim2, dim3) produced from two-dimensional simulations
with a 15$M_\odot$ progenitor star mass and increasing rotational velocity. 

The right panel of the Fig.~\ref{fig:sdc_initial} shows the detection efficiency curves for the Yakunin waveforms for two and single detector networks.  From this plot it can be concluded that the detectability of the waveforms for two and single detector networks are comparable. One interesting observation is that the efficiency goes higher for H1H1 network comparing to H1L1 network, because the injected waveforms between two detectors are fully coherent.

Given Fig.~\ref{fig:sdc_initial} it is clear that a significant challenge with single detector analysis is suppressing the non-linear loud noise transients. It is also important that we remove glitches without decreasing the sensitivity of the algorithm to detect GW signals.

\section{Machine Learning Algorithm}\label{machinelearning}

In this section we introduce the ML algorithm employed in our work. As a full introduction to the method is beyond the scope of this paper, we present only the information that is essential for the understanding of our analysis. For a deeper discussion, the reader is referred to Ref.\ \cite{GPweb}.

The supervised ML algorithm employed is a type of evolutionary computation called genetic programming (GP). It is an analog to biological natural selection. In GP, we evolve a population of programs through successive generations to solve a particular, defined problem \cite{Koza1992}. An individual evolved GP program is an hypothesis which when executed takes the form of a mathematical, multivariate expression.

In the training process each evolved hypothesis is executed against each given data point (sample from the real world) and in turn generates a prediction. Each and every prediction in the training set is compared against the correct, qualified label (truth). The distance between the prediction and truth is used to generate a fitness score for each hypothesis. As the evolutionary process in GP favors those hypotheses with a higher fitness, they are more likely to pass some or all of their code into the next generation of evolved hypotheses. Lesser performing individuals are, over successive generations, abandoned. Therefore, GP programs that demonstrate a higher overall fitness are more likely (but not guaranteed) to be selected for the next generation. Thus, each subsequent generation of programs is more likely (but not guaranteed) to solve the given problem than the prior \cite{Poli2008}.

GP multivariate expressions are classically represented as a syntax tree, where the trees have a root (top center), nodes (mathematical operators), and
leaves (operands). Operators can be arithmetic, trigonometric, and boolean, for example. As with any mathematical expression, operands are variable
place-holders for the real-world values. When evaluated, the real-world data are substituted for the variables in the multivariate expressions, data
point by data point. The depth of a tree determines the complexity of the evolved multivariate expression. Deeper trees are able to incorporate more
operands in each expression, and tend toward non-linear functions.

With GP the user assigns run-time parameters such as the quantity of individual trees in the initial population, type of GP trees employed, the number of individual programs selected for each tournament (a comparison of fitness scores), and the termination criterion (eg: number of generations). The performance of the algorithm can be tuned through the selection of these parameters.

The work-flow of a generational GP run incorporates three basic steps: a) Generation of an initial, stochastic population; b) Iterative selection, evaluation, and application of genetic operations (reproduction, mutation and crossover); c) transfer of the evolved copy into the subsequent generation. Steps b) and c) repeat until the user-defined termination criterion is met \cite{Poli2008}.

In this advanced era of machine learning, many algorithms tend toward black box solutions, both off-line training and on-line processing without a working knowledge of how any given solution was derived. It is important to many researchers, and their fields of research, to understand the internal workings of any system, including computer software.

The GP algorithm employed in this body of research offers total transparency to its internal workings, and the opportunity to review the evolutionary process at each step, pause, archive, and continue. Moreover, as the GP hypothesis is a stand-alone mathematical expression whose variables call upon data features generated outside of GP, it can be readily employed as a portable model for online data classification or regression analysis in any number of shell, script, or compiled computer languages.

In our analysis we used a tree-based open source python code, Karoo GP \cite{KarooGP}, that was originally written by one of the authors (KS) for the mitigation of RFI in radio astronomy at the Square Kilometre Array \cite{StaatsThesis}. Karoo GP is scalable, with multicore and GPU support enabled by the Python library TensorFlow and the capacity to work with very large datasets \cite{Gecco2017}.

\section{Data preparation}

We train the GP algorithm on the families of waveforms described in Sec.\ \ref{cWB} (defined as $Class 1$ for the sake of the GP algorithm) and the background events ($Class 0$). We use 1.47 days of background events sampled from a different time frame in the LIGO/Virgo O1 observing run, else we would otherwise bias the analysis.

Each dataset is built by combining a number of simulations ranging from a few hundreds to a few thousands per model family, depending on the family, and a comparable number of background triggers randomly extracted from the total number of cWB events in the analysis time frame. The dimension of the datasets is therefore determined by the number of available injected simulations, and a similar number of background triggers combined.

Each trigger is identified by a GPS time stamp and an 11-dimensional vector that contains the cWB parameters. In our analysis we employ this data vector with a cWB parameter subet relevant to a single-interferometer configuration. The vector elements are as follows:

\begin{itemize}
\item rho0 - ranking statistic (effectively the signal-to-noise ratio of the event)
\item volume0 - event volume, i.e., the number of wavelet defining the trigger
\item duration0 - energy-weighted duration
\item duration1 - raw duration of the trigger in the time-frequency domain
\item frequency0 - central frequency of the trigger computed from the reconstructed waveform
\item frequency1 - energy-weighted central frequency estimated in the time-frequency domain
\item low0 and high0 - minimum and maximum frequencies associated to the time-frequency map pixels
\item bandwidth0 - energy-weighted bandwidth
\item bandwidth1 - raw bandwidth
\item norm - event's norm factor or ellipticity. 
\end{itemize}

For the sake of trainig the ML algorithm, triggers corresponding to injected CCSN waveforms are labeled 1 (positives), and background events are labeled 0 (negatives). The triggers in the datasets are then randomly shuffled and split into thirds. Two thirds are used for ML training and internal testing. The remaining one third is reserved for external, blind validation. Table \ref{tab:datasets} shows the analyzed datasets and their dimensionality.

\begin{table*}[t]
\begin{center}
{\small %
\begin{tabular}{|l||*{6}{c|}}\hline
\backslashbox{\textbf{Family}}{\textbf{Distance (kpc)}}
&\makebox[3em]{\textbf{1.00}}&\makebox[3em]{\textbf{1.78}}&\makebox[3em]{\textbf{3.16}}&\makebox[3em]{\textbf{4.22}}&\makebox[3em]{\textbf{5.62}}
&\makebox[3em]{\textbf{7.50}}
\\\hline\hline
\textbf{Ott \cite{Ott:2012mr}} & 2000/1000 & 2000/1000 & 2000/1000  & -- & -- & -- \\ \hline
\textbf{Dimmelmeier \cite{Dimmelmeier:2008iq}} & 6000/3000 & 6000/3000 & 6000/3000  & -- & 6000/3000 & -- \\ \hline
\textbf{Scheidegger \cite{Scheidegger:2009va}} & 4000/2000  & 4000/2000 & 4000/2000  & -- & 4000/2000 & 4000/2000 \\ \hline
\textbf{Yakunin \cite{Yakunin:2015wra}} & 2000/1000  & 2000/1000 & 2000/1000  & -- & 3125/1125 & -- \\ \hline
\textbf{All combined} & 5500/2750  & 5500/2750 & 5500/2750  & 5500/2750 & --  & -- \\ \hline
\end{tabular}
}
\caption{Different waveform families and injection distances used in the analysis. The distances are chosen to be within the detector galactic range and
equally spaced in the logarithmic scale. The entries in the table give the dimensions of the full training + validation sets. The first value is the total number of
triggers in the dataset. The second value is the number of simulations. Two thirds of the triggers are used for training and internal testing. The remaining
one-third is used for external validation.}
\label{tab:datasets}
\end{center}
\end{table*}

\section{Analysis and Results}

With evolutionary computation, for which GP is a subset, the parameters employed by a given algorithm are explored to avoid local minima and then
optimized for incremental improvements and the speed at which the algorithm arrives to the desired solution. We train the GP algorithm with a population of 300
individuals, 100 generations, tournament size set to 20, and a max (min) tree depth of 4 (3). These values are in line with well-established choices for the use of
GP on datasets of similar dimensionality to the datasets considered here \cite{Cavaglia:2018xjq,Poli2008}. This optimal combination was tested by varying the
population size, number of individuals selected for the tournament, number of generations and tree depth in preliminary runs. To minimize the uncertainty in the
determination of the multivariate expressions used for the classification of the triggers inherent in the stochastic nature of GP process, we conduct a full
evaluation 200 times, on each dataset.

We first use the dataset with Dimmelmeier waveforms injected at a distance of 3.16\,kpc, and discuss the results in detail. We then compare these results with those obtained with other injected models, at various distances and mixed datasets.

\begin{figure*}[t]
  \begin{center}
    \begin{tabular}{  l  l  }
    \includegraphics[width=80mm]{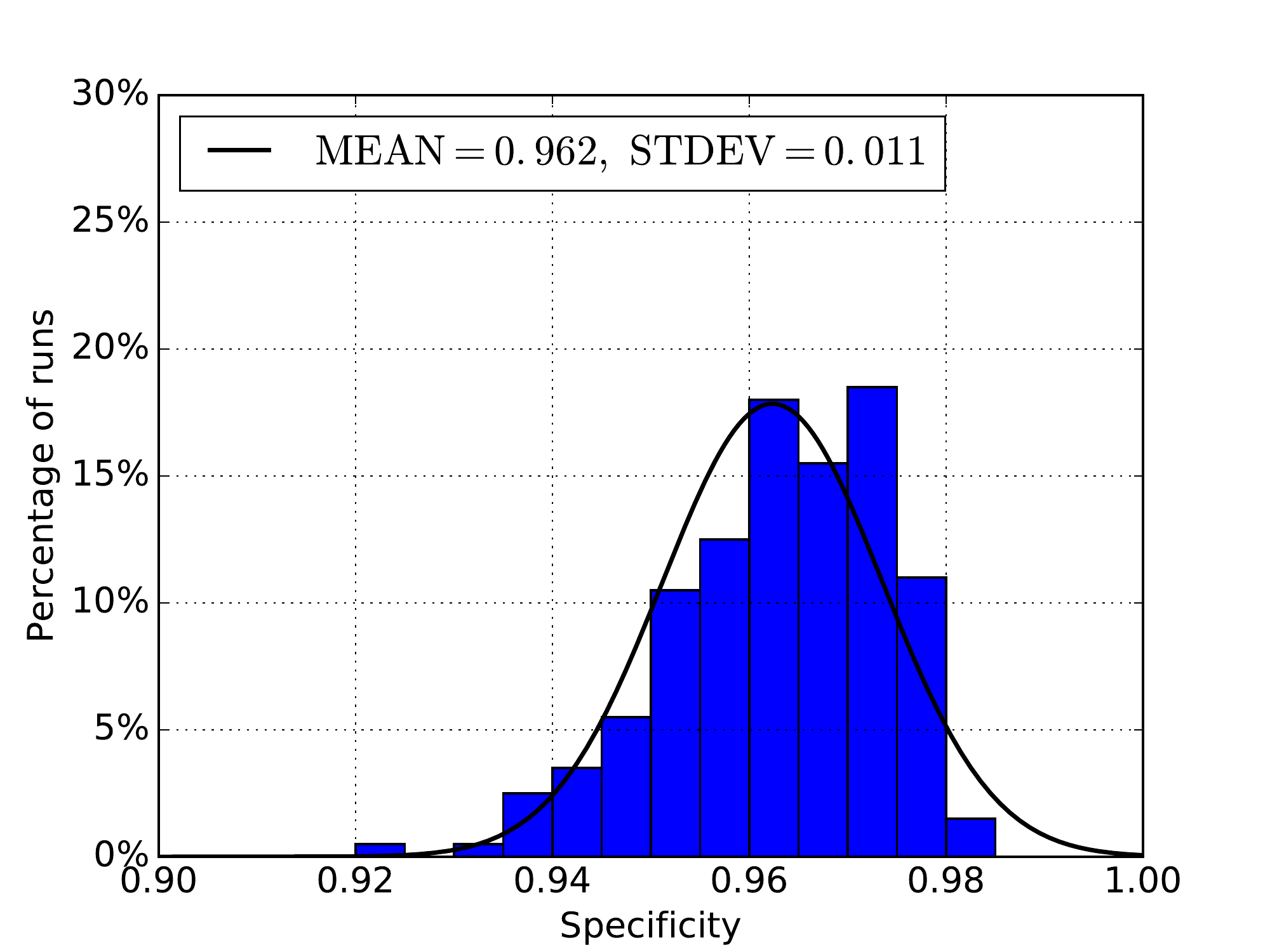}&
    \includegraphics[width=80mm]{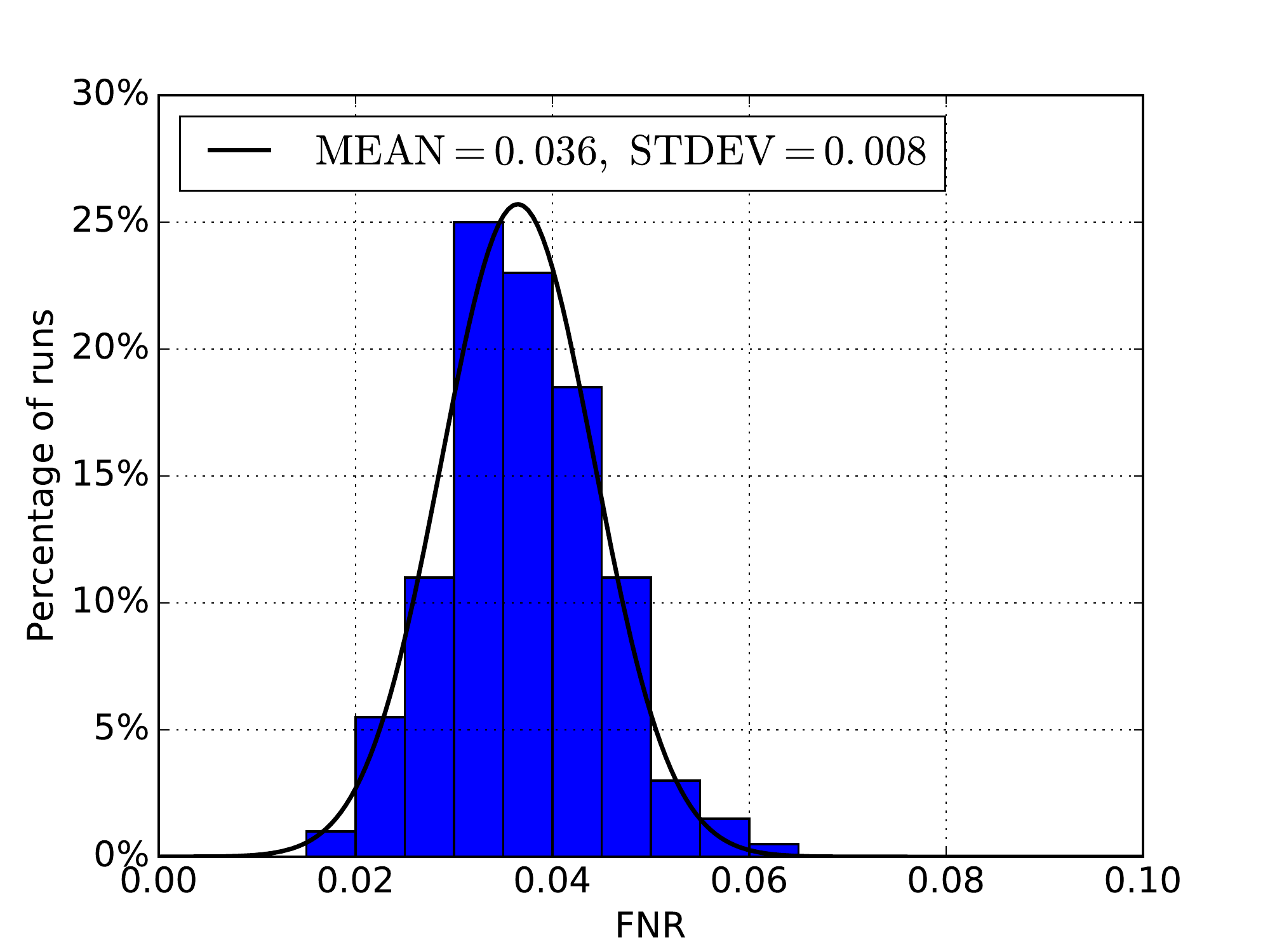}\\
   \end{tabular}
  \end{center}
  \caption{Specificity (left panel) and FNR (right panel) for the testing dataset with Dimmelmeier injected waveforms at 3.16\,kpc. The histograms represent the results for 200 runs}\label{fig:Dim123-1}
\end{figure*}

As the goal of the analysis is to reduce the search background, the relevant quantities in the confusion matrix are the specificity (True Negative Rate, TNR) and the False Negative Rate (FNR), i.e., the number of signals mistakenly identified as noise. Figure \ref{fig:Dim123-1} shows specificity and FNR for the Dimmelmeier dataset with waveforms injected at 3.16\,kpc. The GP algorithm is able to identify on average 96.2\% of the noise transients while misclassifying on average only 3.6\% of GW signals. Even for the worst runs, the number of lost signals remains well below 1\% with a glitch removal efficiency above 92\%.

\begin{figure*}[t]
  \begin{center}
    \begin{tabular}{  l  l  }
    \includegraphics[width=80mm]{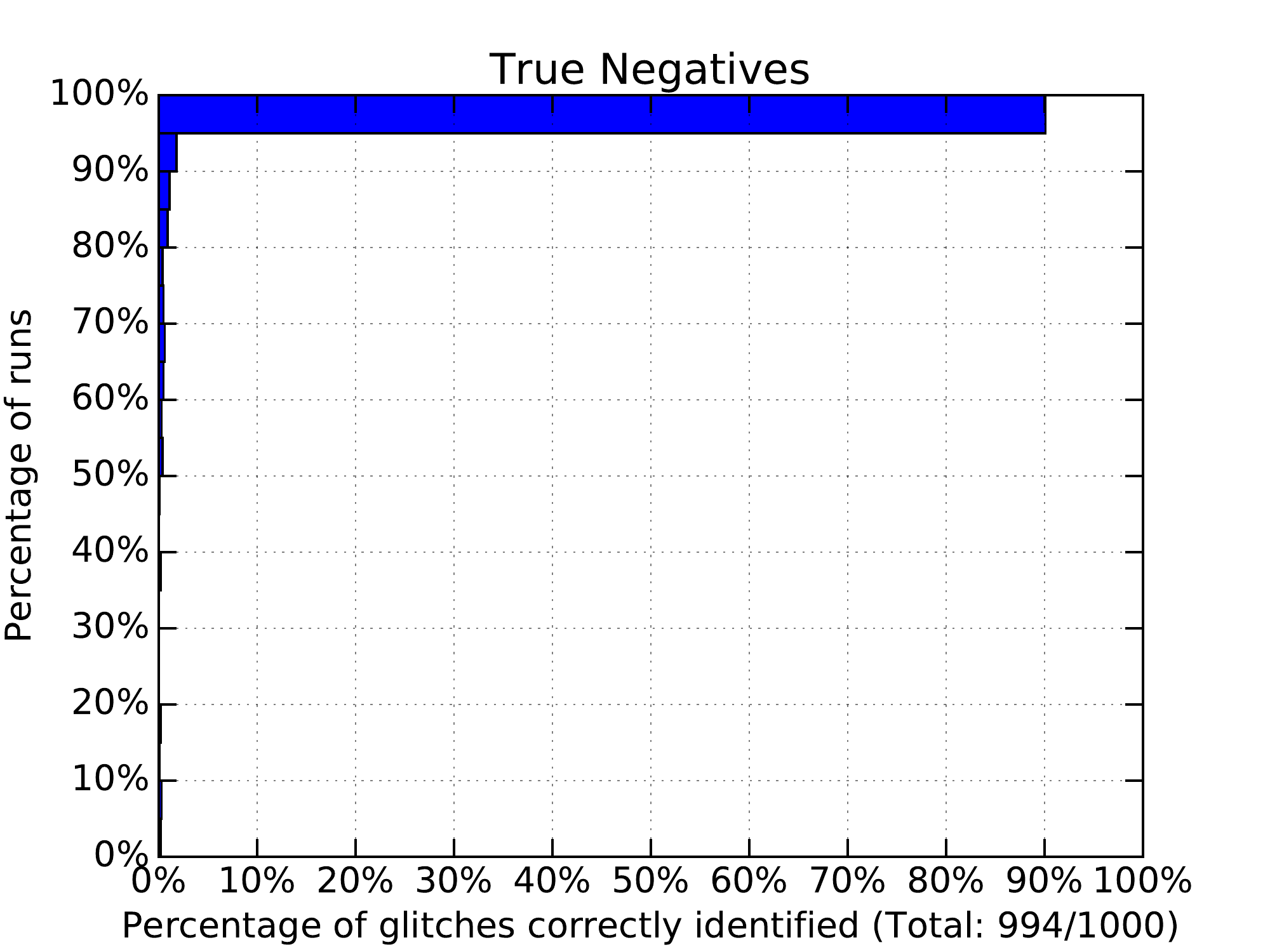}&
    \includegraphics[width=80mm]{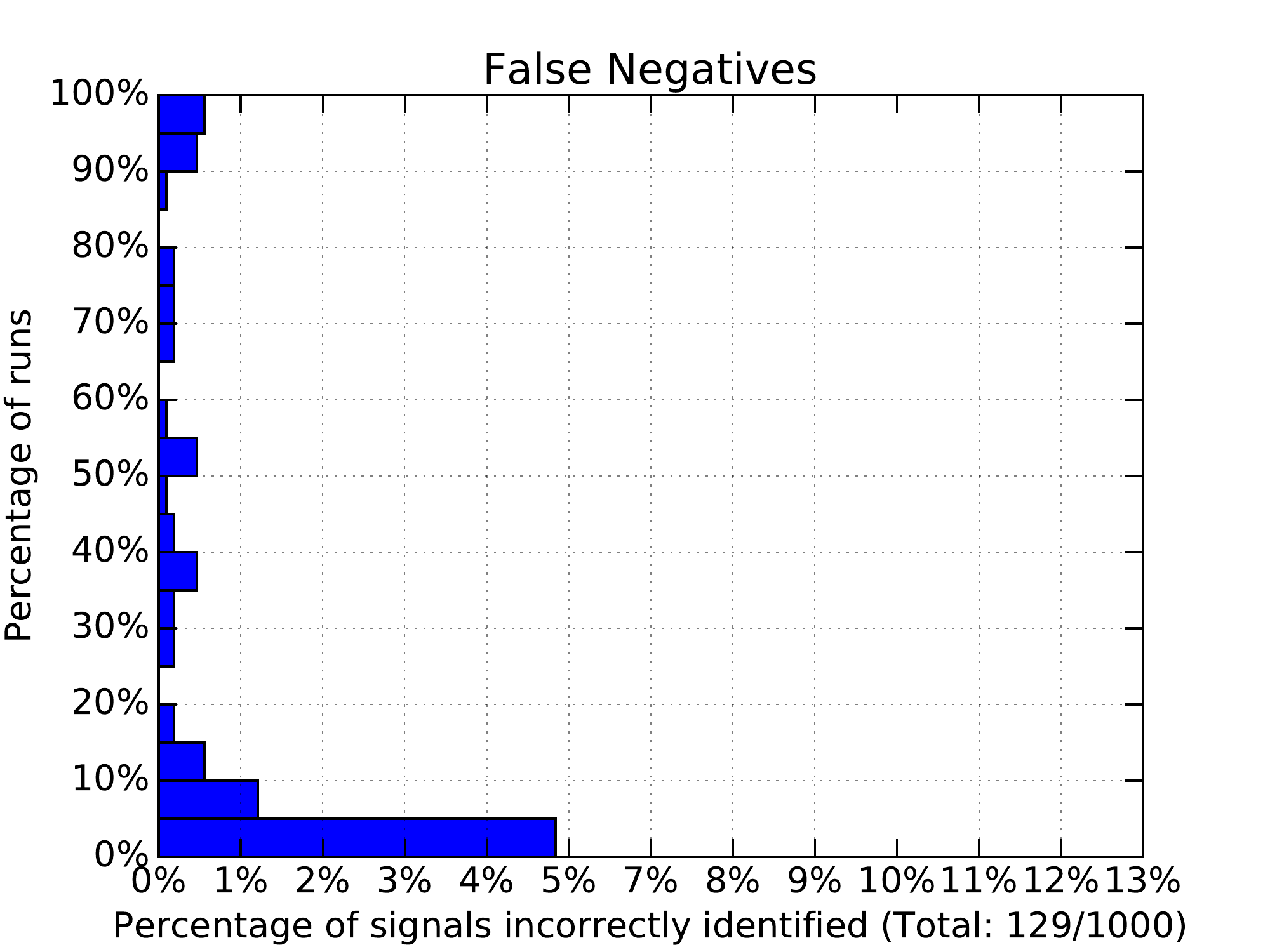}\\
   \end{tabular}
  \end{center}
  \caption{Percentage of runs that correctly identify glitches (true negatives, left panel) and mistakenly identify signals (false negatives, right panel) as a function of the percentage of triggers for the testing dataset with Dimmelmeier injected waveforms at 3.16\,kpc. Top panel: About 90\% of the glitches are correctly identified by 95\% or more of the runs. Bottom panel: Less than 1\% of the signals are misclassified by 95\% or more of the runs.}\label{fig:Dim123-2}
\end{figure*}

\begin{figure*}[t]
  \begin{center}
      \includegraphics[width=170mm]{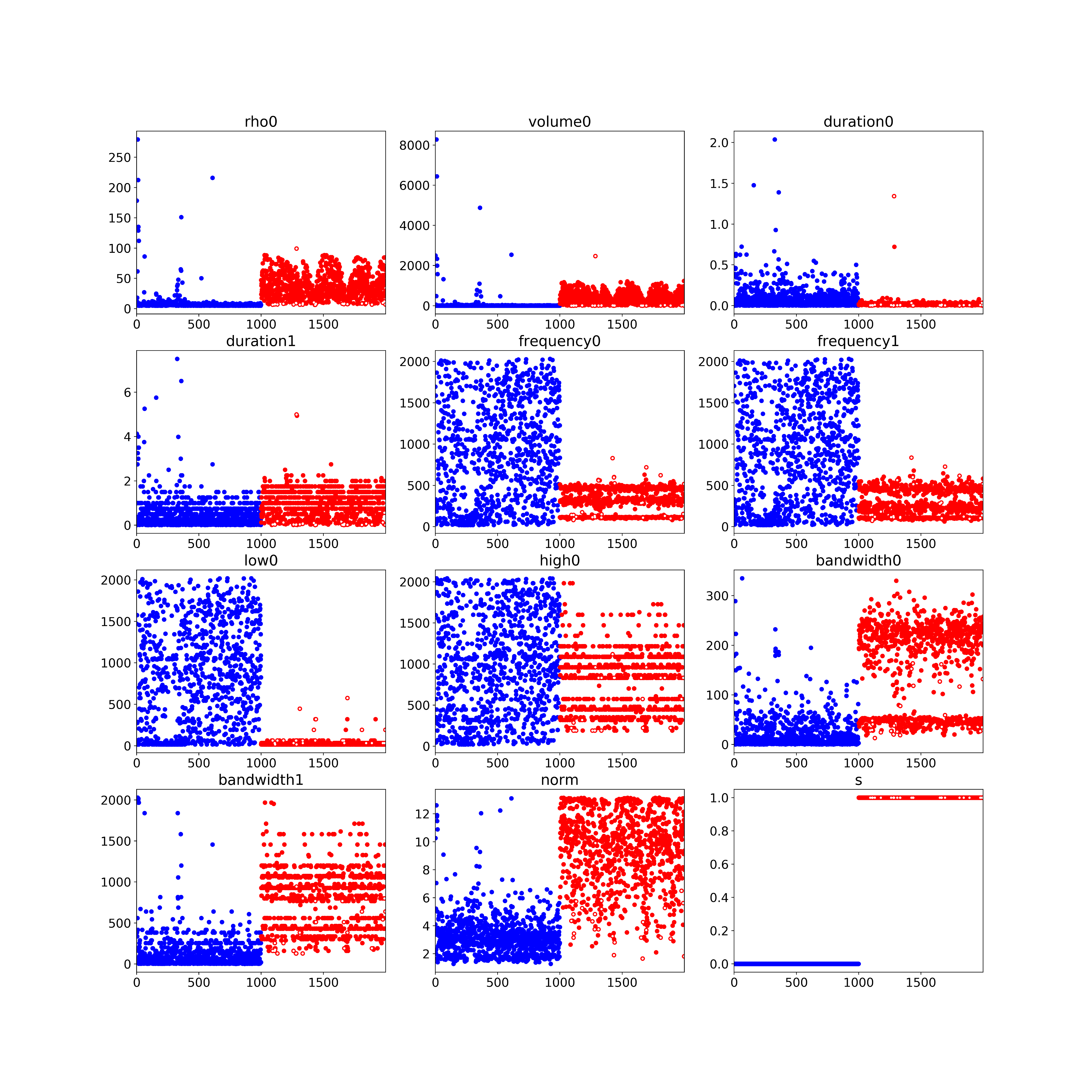}
  \end{center}
  \caption{Plots of false negatives for the testing dataset with Dimmelmeier injected waveforms at 3.16\,kpc. The dataset contains 1000 injections (red full circles) and 1000 noise triggers (blue full circles). Different panels show the distribution of the cWB parameters (ML features) across triggers with the bottom-right panel showing the trigger label. The $x$ axis of the panels denotes the index of the trigger, the $y$ axis gives the value of the corresponding cWB parameter. When a threshold for trigger identification of 60\% on the number of runs is applied, 25 injections (2.5\%) are misclassified (red empty circles).}\label{fig:Dim123-3}
\end{figure*}

The performance of the runs as a function of the specificity and FNR metrics is shown in Fig.\ \ref{fig:Dim123-2}. The top (bottom) panel shows the percentage of glitches (signals) correctly (incorrectly)
identified by a given percentage of runs (in bins of 5\%). The majority of runs correctly identifies the noise transients while misidentifying only a small percentage of signals: About 90\% of the glitches
are correctly identified by 95\% or more of the runs while less than 1\% of the signals are misidentified as noise by 95\% of more of the runs. Even if the threshold on the number of runs is
reduced to 60\%, we can still correctly identify over 95\% of the noise transients while losing only about 3\% of the signals (see Fig.\ \ref{fig:Dim123-3}).

\begin{figure}[htbp]
  \begin{center}
      \includegraphics[width=80mm]{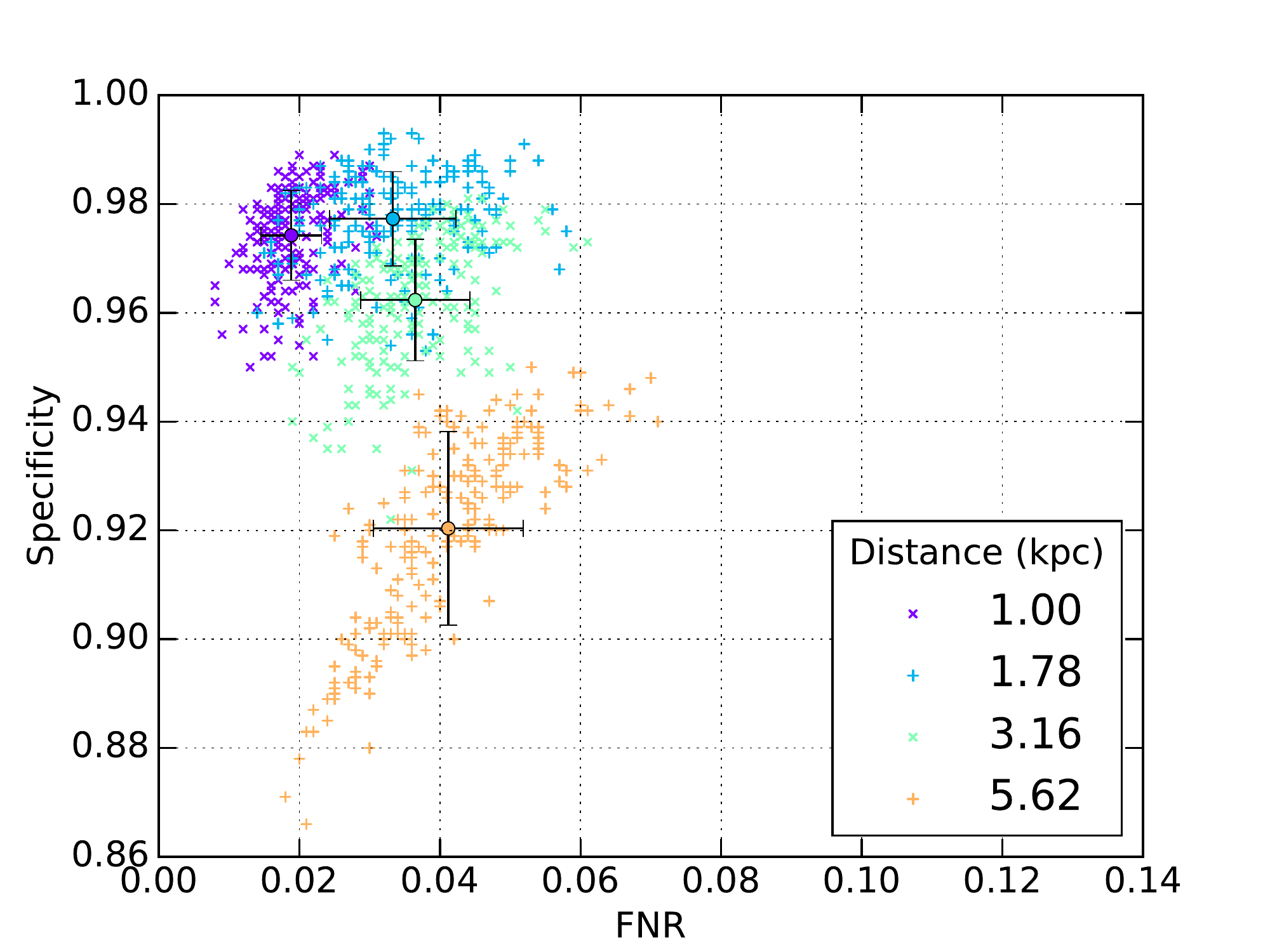}
  \end{center}
  \caption{Scatterplot of specificity vs.\ FNR for datasets with Dimmelmeier waveforms injected at different distances (1.00, 1.78, 3.16 and 5.62\,kpc). Each point represents a GP run (200 runs total for each dataset). Average values with standard deviations are also shown.}\label{fig:Dim123-4}
\end{figure}

The performance of the GP classification varies across datasets as the distance of the injected waveforms varies. The farther the distance of the simulated GW, the smaller is its SNR. Thus it is more difficult to distinguish injections from noise triggers. Figure \ref{fig:Dim123-4} shows how the specificity and FNR vary as a function of injection distance for the Dimmelmeier waveforms. While the performance of the classification diminishes as the injected distance increases, even for the largest injection distance tested, 5.62\,kpc, the average specificity across the run remains above 92\% with FNR below 4\%.

\begin{figure}[htbp]
  \begin{center}
      \includegraphics[width=80mm]{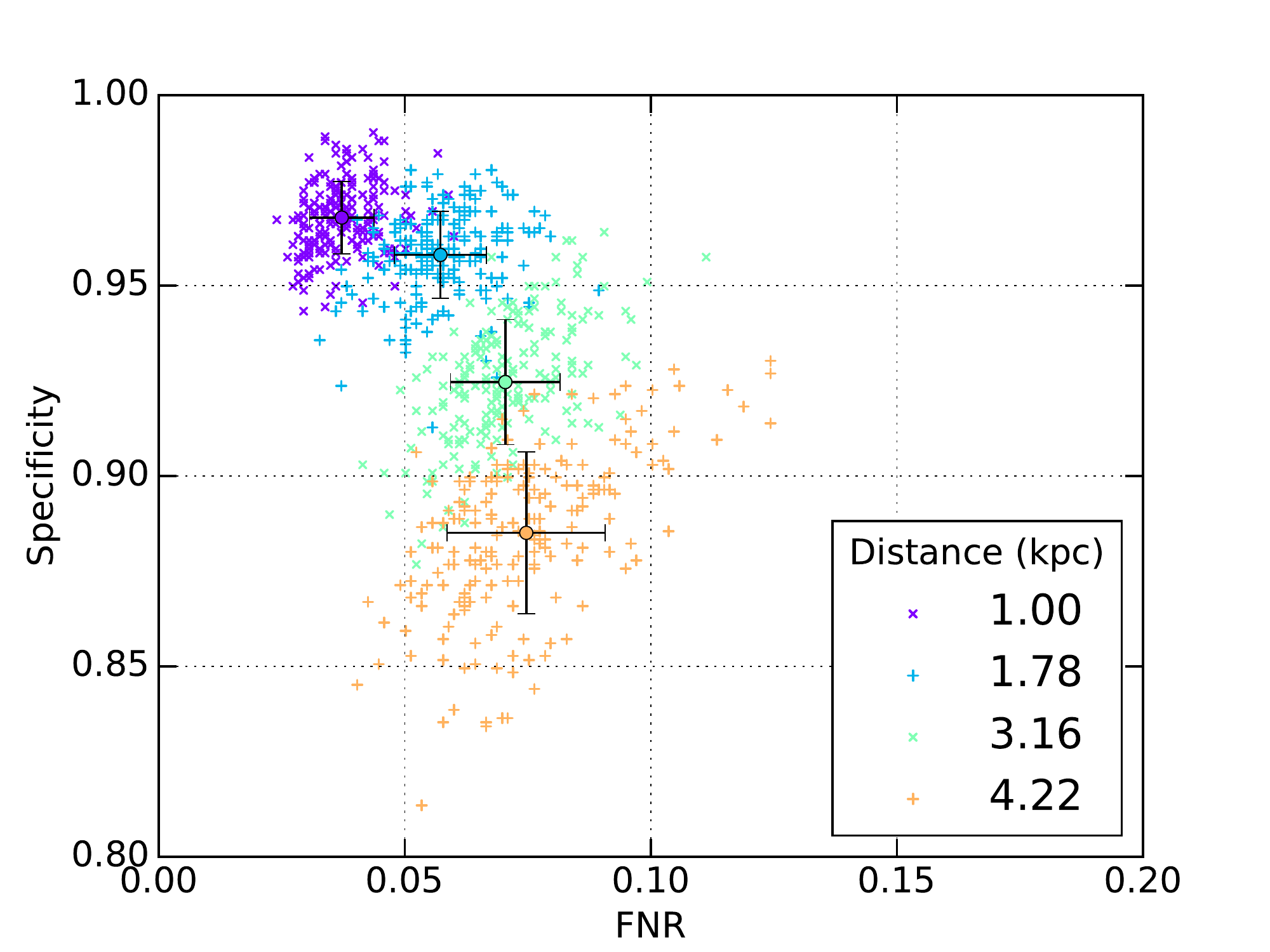}
  \end{center}
  \caption{Scatterplot of specificity vs.\ FNR for all combined waveforms (Ott, Dimmelmeier, Scheiddeger and Yakunin) injected at different distances (1.00, 1.78, 3.16 and 4.22\,kpc). Each point represents a GP run (200 runs total for each dataset). Average values with standard deviations are also shown.}\label{fig:Allcombined-1}
\end{figure}

We repeat the analysis above for all other datasets in table \ref{tab:datasets} with overall similar results. Scheidegger waveform models seems to fare better
with average values of specificity (FNR) decreasing from over 99\% (less than 1\%) at 1.00\,kpc to about 97\% (3\%) at 7.5\,kpc. Ott and Yakunin models
typically do worse than Dimmelmeier models with average specificity (FNR) ranging from about 97\% (5\%) and 98\% (5\%) at 1.0\,kpc to about 86\% (12\%) and
93\% (9\%) at 3.16\,kpc for Ott and Yakunin waveforms, respectively. This may be due to the Dimmelmeier waveforms being more energetic and compact in the
time-frequency space than the Yakunin waveforms models. As the physics of CCSN is uncertain, we also trained the GP algorithm on ``agnostic'' datasets by
combining waveforms from all different models. The classification performance of the GP algorithm remains comparable to the performance of the single-model
training. Figure \ref{fig:Allcombined-1} shows specificity vs. FNR for the combined datasets with injected waveforms at distances from 1.0 to 4.22\,kpc. Even
for the largest distances, the average specificity remains above about 88\% with FNR less than about 8\%. We conclude that the procedure is robust against the
different CCSN physical models. 

\begin{figure*}[t]
  \begin{center}
    \begin{tabular}{  l  l  }
      \includegraphics[width=70mm]{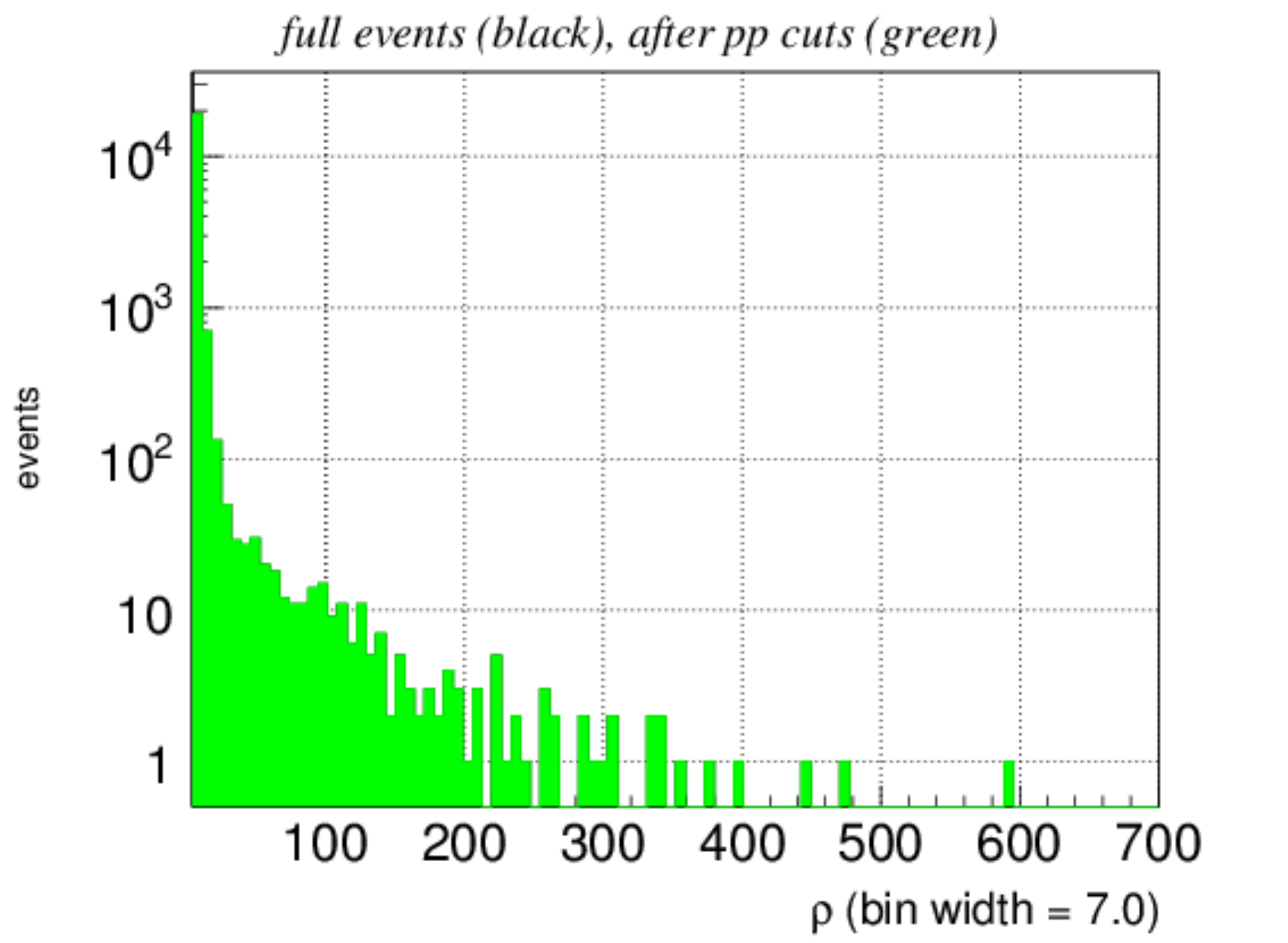}&  \includegraphics[width=70mm]{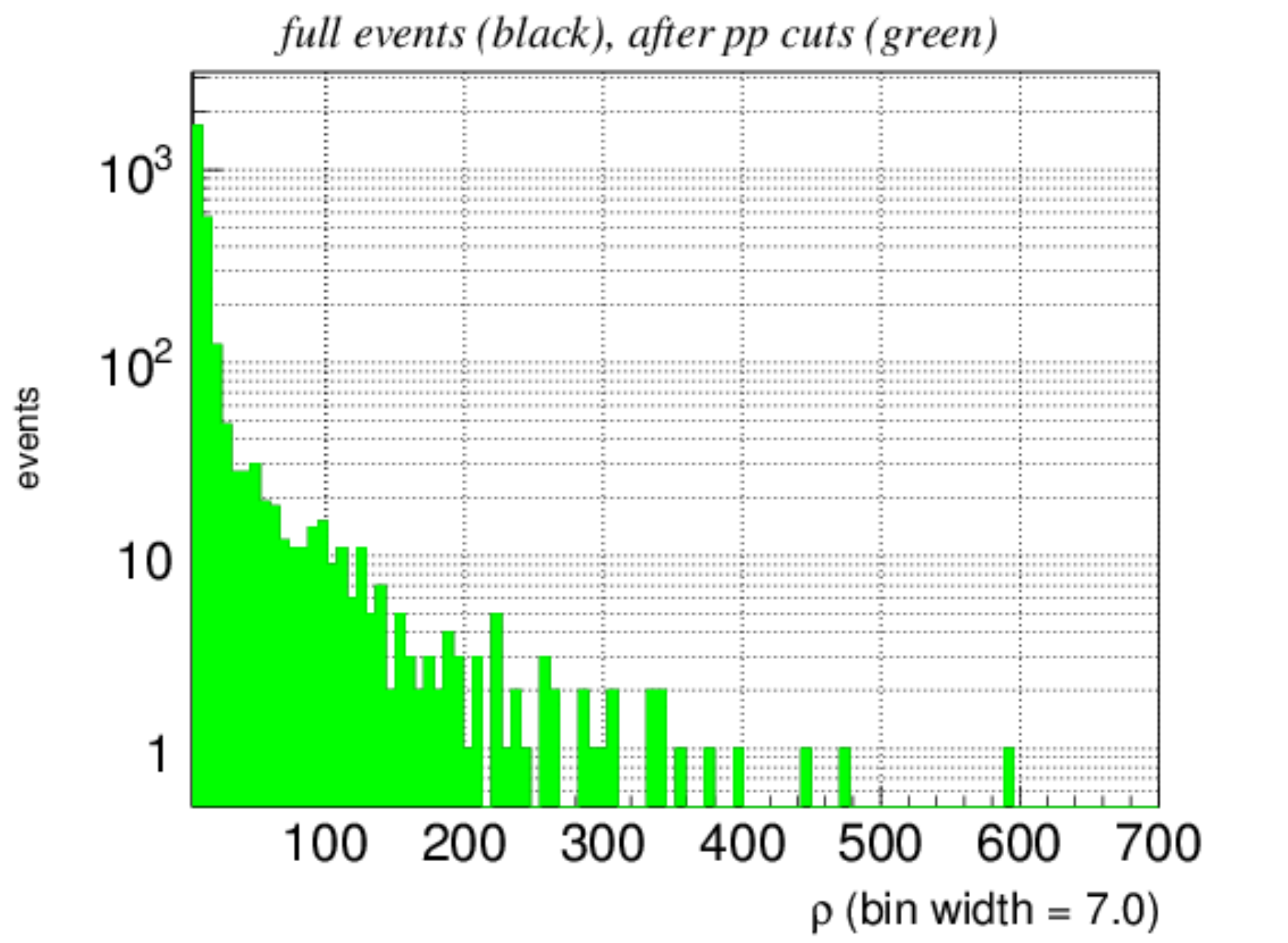}
   \end{tabular}
  \end{center}
  \caption{cWB background on a two-day period before (left) and after (right) the cleaning procedure is applied. The GP algorithm to produce the clean background has been trained with all waveform models injected at a distance of 3.16\,kpc.}
\label{fig:cleaning1}
\end{figure*}

Once the GP algorithm has been trained, it can be used to reduce the cWB search background.  We first classify the cWB triggers and then remove the triggers that are identified as noise transients by 90\% of the GP runs. Figure \ref{fig:cleaning1} shows the cWB background for the two-day period before and after the cleaning procedure, where for the latter we have used the training obtained with all waveform models at 3.16kpc. (Other training sets give comparable results.) The number of triggers in the background is significantly removed, specifically at low SNR where the number of triggers is decreased by a factor $\sim 10$, as expected.

\begin{figure}[htbp]
  \begin{center}
      \includegraphics[width=70mm]{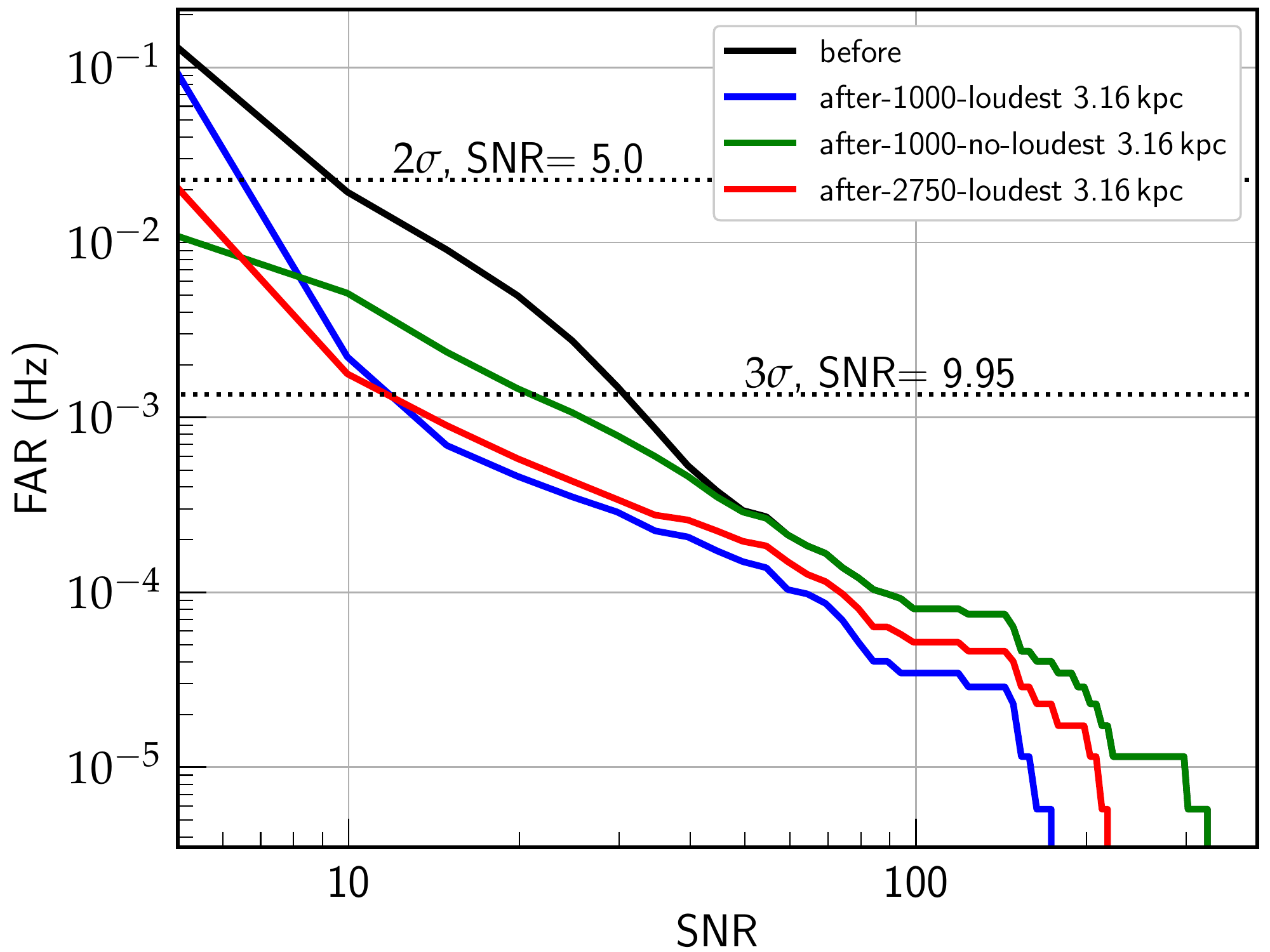}
  \end{center}
  \caption{cWB background on a two-day period before and after the cleaning procedure is applied. The black curve shows the original background. The green, blue and red curves show the background after applying the results of the training obtained on a dataset with all waveform models injected at 3.16\,kpc and random background selection, 1000 loudest background triggers and 2750 background triggers, respectively. }\label{fig:cleaning2}
\end{figure}

As we are mostly interested in removing loud triggers to increase the detection confidence level, we can bias the GP algorithm to remove high-SNR triggers by training it on a dataset that includes only a subset of the loudest triggers. The results for the background on a (different) O1 period are shown in Fig.\ \ref{fig:cleaning2}. By biasing the training dataset towards high-SNR triggers (blue and green curves), the search background is not cleaned as efficiently at low SNR as the background cleaned by training on a dataset with a random selection of background triggers. However, the high-SNR tail of the background shows a reduction by more than one order of magnitude with significant gains down to about SNR$\sim 10$. The cleaning procedure lowers the SNR required for 3$\sigma$ c.l. detections by a factor up to $\sim 3$ and $\sim 2$, respectively.

Table~\ref{tab:cwb_sim} presents the impact of ML noise removal on the detection efficiency.
We show how the detection efficiency before and after changes after ML is applied together with a percent error decrease. On average, the decrease in the detection efficiency for $2\sigma$ c.l. is about 10\,\%, while for $3\sigma$ c.l. it is about 30\,\%.

\begin{table*}[t]
\begin{center}
\begin{tabular}{c|c|c}
\hline
\hline
Waveform & $2\sigma$ c.l. & $3\sigma$ c.l. \\ 
\hline
\hline
ott1 & 0.224 / 0.1946 (13.13\%) & 0.060 / 0.042 (30.00\%) \\ \hline
yak1 & 0.379 / 0.339 (10.55\%) & 0.286 / 0.200 (30.07\%) \\ \hline
yak2 & 0.506 / 0.452 (10.67\%) & 0.437 / 0.307 (29.75\%) \\ \hline
yak3 & 0.52 / 0.465 (10.58\%) & 0.455 / 0.319 (29.89\%) \\ \hline
yak4 & 0.66 / 0.590 (10.61\%) & 0.610 / 0.428 (29.84\%) \\ \hline
sch2 & 0.977 / 0.874 (10.54\%) & 0.974 / 0.684 (29.77\%) \\ \hline
sch3 & 0.988 / 0.884 (10.53\%) & 0.982 / 0.690 (29.74\%) \\ \hline
dim1 & 0.761 / 0.681 (10.51\%) & 0.693 / 0.487 (29.73\%) \\ \hline
dim2 & 0.835 / 0.747 (10.54\%) & 0.808 / 0.568 (29.70\%) \\ \hline
dim3 & 0.911 / 0.815 (10.54\%) & 0.872 / 0.613 (29.70\%) \\ 
\hline
\hline
\end{tabular}
\caption{Impact on the detection efficiency for the waveforms injected at 3.16\,kpc. Each cell shows the detection efficiency before and after ML is applied with the percent error.}
\label{tab:cwb_sim}
\end{center}
\end{table*}

\begin{figure}[htbp]
  \begin{center}
      \includegraphics[width=70mm]{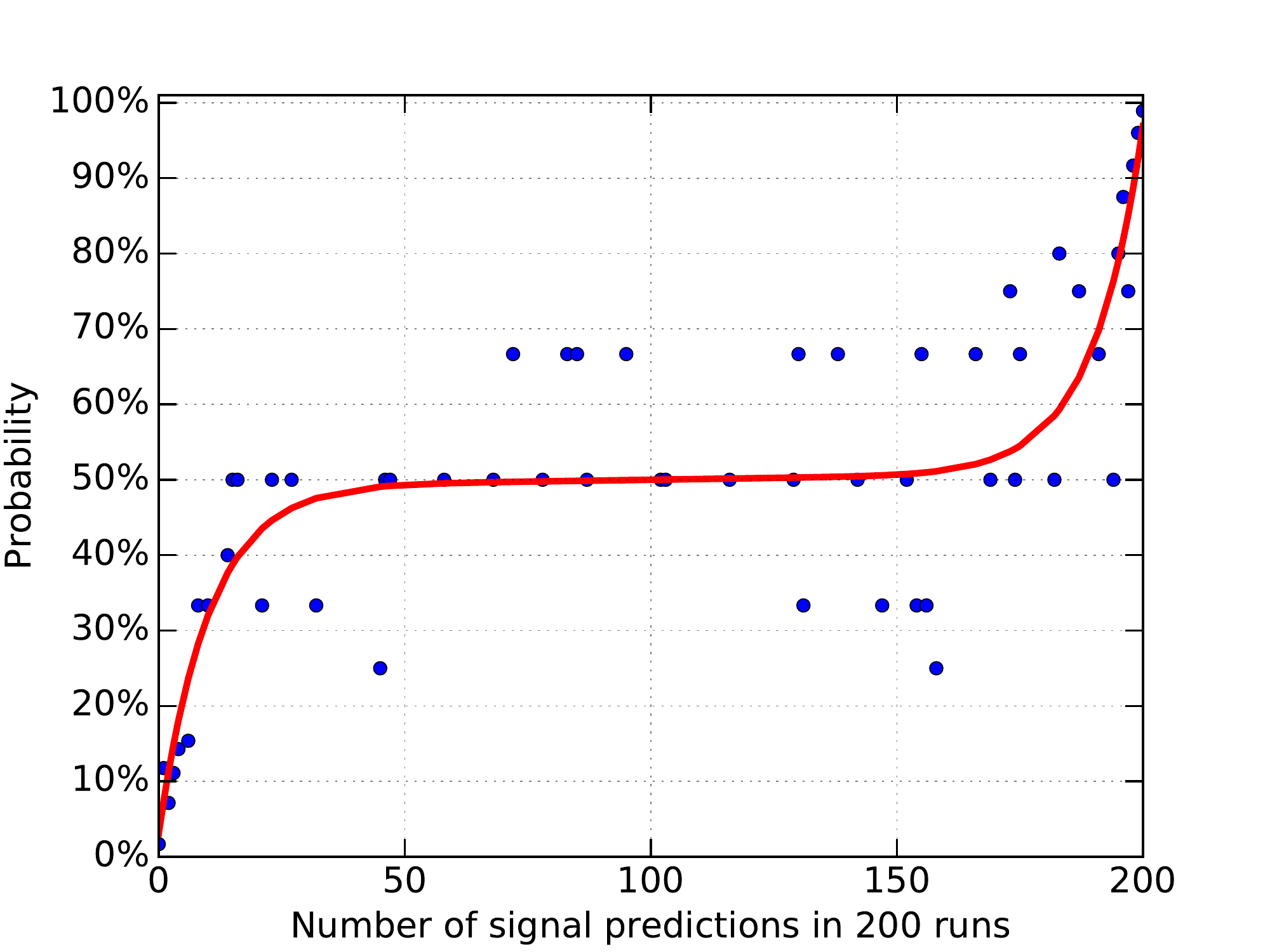}
  \end{center}
  \caption{Likelihood of a trigger being a signal as a function of the number positive identifications. The blue markers represent the likelyhood values obtained from the all-combined training dataset with injected waveforms at 3.16\,kpc. The red curve gives a polynomial best-fit.}\label{fig:Allcombined-prob}
\end{figure}

As a bonus to background reduction, the GP training can also be used to assign a probability to a search trigger being a signal or background noise. According to Bayes theorem, the likelihood that a trigger is a signal $(s)$ if it is classified as such by $n_+$ trained multivariate expressions is
\begin{equation}
P(s|n_+)=\frac{P(n_+|s)P(s)}{P(n_+)}\,,
\label{bayes}
\end{equation}
where $P(n_+|s)$ is the likelihood of observing $n_+$ runs given a signal, $P(s)$ is the probability of observing a signal, and $P(n_+)$ is the probability of
observing $n_+$ GP runs. Using the testing dataset, we estimate these quantities as $P(n_+|s)=n_{TP}(n_+)/n_{s}$, $P(s)=n_{s}/n_T$ and
$P(n_+)=[n_{TP}(n_+)+n_{FP}(n_+)]/n_T$, where $n_{s}$ is the number of signals in the testing dataset containing $n_T$ total triggers ($n_{s}$ signals + $n_{b}$
background), and $n_{TP}(n_+)$ [$n_{FP}(n_+)$] is the number of triggers in the dataset positively [mistakenly] identified $n_+$ times, respectively. The likelihood that a
trigger is a signal given $n_+$ positive identifications is then
\begin{equation}
P(s|n_+)=\frac{n_{TP}(n_+)}{n_{TP}(n_+)+n_{FP}(n_+)}\,.
\label{probability}
\end{equation}
The likelihood of a trigger being a signal is shown in Fig.\ \ref{fig:Allcombined-prob} for the training on all combined CCSN waveforms injected at 3.16\,kpc. The plot shows the likelihood values (blue markers) with a polynomial best fit (red curve). For this particular training, the probability of a trigger to be a signal when it is identified by 200 (190) runs is roughly 96\% (68\%). 

As the detection of CCSN signals is very complex and only a limited number of simulated waveforms are available, in order to evaluate possible bias factors in the procedure, we tested the
algorithm by classifying a set of waveforms with multivariate expressions obtained by training on a different set of waveforms on a different epoch. The different characteristics of the noise and waveforms
in the training and testing sets can be used as a proxy for the stochastic nature and the unknown physical features of the waveforms in a real case. We applied the method to a set of 18 triggers (including
simulated signals and background) in a blind analysis. Table~\ref{tab:blind} shows for each trigger the number of runs which identify the trigger as a signal and the probability value $P(s|n_+)$ obtained
with the training on the dataset with all-combined waveforms modes injected at 3.16\,kpc. Seven background triggers (70\%) are correctly identified as background with $97$\% and six injected signals (one
at 3.16\,kpc and five at 1\,kpc) are correctly identified as signals with probability $>62$\%. Only for three triggers (one background trigger, 1137088400.326843, and two simulated injections
1137123606.447540 and 1137250081.748009, the probabilities are close to 50\% and their identification as background or signal may be considered inconclusive.

\begin{table}[htbp]
\begin{center}
\begin{tabular}{l@{\hspace{10pt}}|@{\hspace{1pt}}r@{\hspace{1pt}}|c|l}
\hline
\hline
Trigger time & $n_+$ & $P(s|n_+)$ & Actual \\ 
\hline
\hline
1137221362.849899 & 0 & 0.03 & BKG\\
1137221296.450439 & 12 & 0.35 &  BKG\\
1137221270.478584 & 7 & 0.26 &  BKG\\
1137221270.315765 & 0 & 0.03 &  BKG\\
1137221256.461151 & 0  & 0.03 &  BKG\\
1137221254.992889 & 0  & 0.03 &  BKG\\
1137221206.790939 & 0  & 0.03 &  BKG\\
1137221187.891924 & 0  & 0.03 &  BKG\\
1137088411.819580 & 0  & 0.03 &  BKG\\
1137088400.326843 & 91  & 0.50 &  BKG\\
1137123606.447540 & 146  & 0.50 &  SIG (Yak, 3.16\,kpc)\\
1137234559.739685 & 188  & 0.65 &  SIG (Yak, 3.16\,kpc)\\
1137250081.748009 & 167  & 0.52 &  SIG (Yak, 3.16\,kpc)\\
1137215815.308205 & 188  & 0.65 &  SIG (Yak, 1\,kpc)\\
1137240747.519287 & 188  & 0.65 &  SIG (Yak, 1\,kpc)\\
1137251495.131439 & 188  & 0.65 &  SIG (Yak, 1\,kpc)\\
1137232392.167053 & 188  & 0.65 &  SIG (Yak, 1\,kpc)\\
1137237558.365189 & 186  & 0.62 &  SIG (Yak, 1\,kpc)\\ 
\hline
\hline
\end{tabular}
\caption{Results of the classification obtained from the training on the dataset with all-combined waveforms modes injected at 3.16\,kpc on a blind set of 18 triggers. The first column gives the GPS time of the trigger, the second column the number of positive (siganl) identifications out of 200 GP runs, the third column gives the probability $P(s|n_+)$ of the trigger being a signal, the last colum gives the actual nature of the trigger [BKG = background, SIG = injected signal (waveform family, distance)]}
\label{tab:blind}
\end{center}
\end{table}

\section{Conclusions}

We presented a new method to reduce the background of LIGO-Virgo searches for GW signals from Galactic CCSN when the detector network is in single interferometer
observing mode. This method consists in  applying a supervised GP ML algorithm to the output of the cWB pipeline. The ML algorithm is trained on datasets of CCSN
waveform simulations and noise background events to classify cWB triggers and remove  events based on their probability of being non-astrophysical. The outcome of
the procedure is an increased statistical significance of GW candidate triggers and a higher detection confidence.

We tested the method on a variety of datasets with different CCSN waveform models injected at fixed Galactic-scale distances. Roughly 90\% or more of
non-astrophysical triggers can be removed from the search  background with a false negative rate of just a few percent, irrespective of the waveform model, even when
the algorithm is trained on datasets with mixed waveforms. To confirm these results we applied the  method on a blind set of triggers and showed that the algorithm
can successfully discriminate noise from simulated GW signals without any prior knowledge of the signal waveform model or injection distance.  The algorithm can be
tuned to enhance specific aspects of the search by introducing a bias during the training process. We illustrated this process by overpopulating the training set
with high-SNR triggers.  This biased dataset allows us to obtain a reduction of over one order of magnitude in the high-SNR tail of the background, which is the most
relevant in case of a detection. The SNR required for a detection with a 3$\sigma$ confidence level is lowered by a factor of $\sim 3$.  Moreover, we
expect these results to improve as more, and more accurate simulations become available in the literature and the algorithm may be trained on a larger pool of GW
waveforms. Although in the case of a real detection of CCSN the GW signal is unlikely to match any of the existing simulations because of the stochastic nature of
the process, the algorithm can be trained to recognize the common physical features of the explosion mechanism by injecting the waveforms on multiple realizations of
the signal noise. 

If applied to current LIGO-Virgo CCSN searches, our method could significantly improve the confidence level of a detection occurring at a time when only a single
interferometer is in observing mode. Our ML algorithm integrates with the cWB pipeline and can be easily trained on any CCSN waveform model or interferometric data.
It would also be straightforward to apply it to a multi-interferometer configuration by  including in the input dataset the full output of a coherent cWB search, as
well as extend it with the inclusion of environmental and instrumental auxiliary channel data from the myriad of interferometric  sensors monitoring the status of
the detectors.

\section*{Acknowledgements}

This work has been partially supported by NSF grants PHY-1921006, PHY-1707668 and PHY-1404139. The authors would like to thank their LIGO Scientific Collaboration and Virgo Collaboration colleagues for their help and useful comments. This work has made use of computational resources provided by the LIGO Laboratory and supported by NSF grants PHY-0757058 and PHY-0823459 and data, software and/or web tools obtained from the Gravitational Wave Open Science Center (https://www.gw-openscience.org), a service of LIGO Laboratory, the LIGO Scientific Collaboration and the Virgo Collaboration. LIGO is funded by the U.S.\ National Science Foundation. Virgo is funded by the French Centre National de Recherche Scientifique (CNRS), the Italian Istituto Nazionale della Fisica Nucleare (INFN) and the Dutch Nikhef, with contributions by Polish and Hungarian institutes. The data that support the findings of this study are openly available.


\bibliography{bibliography}
\end{document}